\pdfoutput=1
\documentclass[11pt]{article}
\usepackage[preprint]{acl}
\usepackage{times}
\usepackage{latexsym}
\usepackage{amsmath}
\usepackage[T1]{fontenc}
\usepackage[utf8]{inputenc}
\usepackage{microtype}
\usepackage{inconsolata}
\usepackage{graphicx}
\usepackage{booktabs}
\usepackage{tabularx}
\usepackage{multirow}
\usepackage{float}
\usepackage{tikz}
\usetikzlibrary{arrows.meta,positioning,fit,backgrounds}

\title{Finbot: From Payment Ambiguity\\to Evidence-Grounded Answers}

\author{Sandip Nandi \and Apurba Bose \and Aishwarya Bose \and Soovadeep Bakshi \\
  Independent Researchers}

\begin{document}
\maketitle

\begin{abstract}
Payment-support conversations combine incomplete descriptions with questions about payment
outcomes, decline reasons and policies. We present Finbot, an open-source, read-only assistant
whose architecture preserves record identity and the financial role of each claim across
clarification, retrieval, explanation, verification and recovery. Its conversation view handles
uncertain details, conflicting selections and corrections; a complementary inspector covers
declines, unknown outcomes, record comparisons and cited policies. Reviewers inspect
claim-to-field traces and inject faults to observe repair, fallback or refusal. In a simulated
clarification component study with eight candidate records, confirmation keeps wrong-record
selections at 2--4\%, versus 21--26\% when acting on the top guess. In 60 authored development
tests, financial-role checks flag every value-preserving corruption, versus 12 with record
binding alone. In two small controlled audits, quoting the cited policy after a failed repair
cuts refusals on 30 live Llama cases from 13 to 4 with no wrong answer under two labellers, and
on 60 public-question tasks, unblinded AI labels rise from 23/60 to 50/60 complete after
requested-field coverage (blind co-author, post-fix: 56/60). The
checks remain conservative: half of correct live drafts need repair or fallback. Code, traces and
cached responses are released.
\end{abstract}

\section{Introduction}

A customer asks ``How much was refunded?'' about one of four \$200 payments. An assistant that
guesses the record, or replies ``Your payment was \$200'', sounds correct while answering the
wrong question about possibly the wrong money. Payment support must resolve incomplete
descriptions, preserve record identity through corrections, and ground each answer in the field
the customer asked about.

\begin{figure*}[t]
\centering
\input{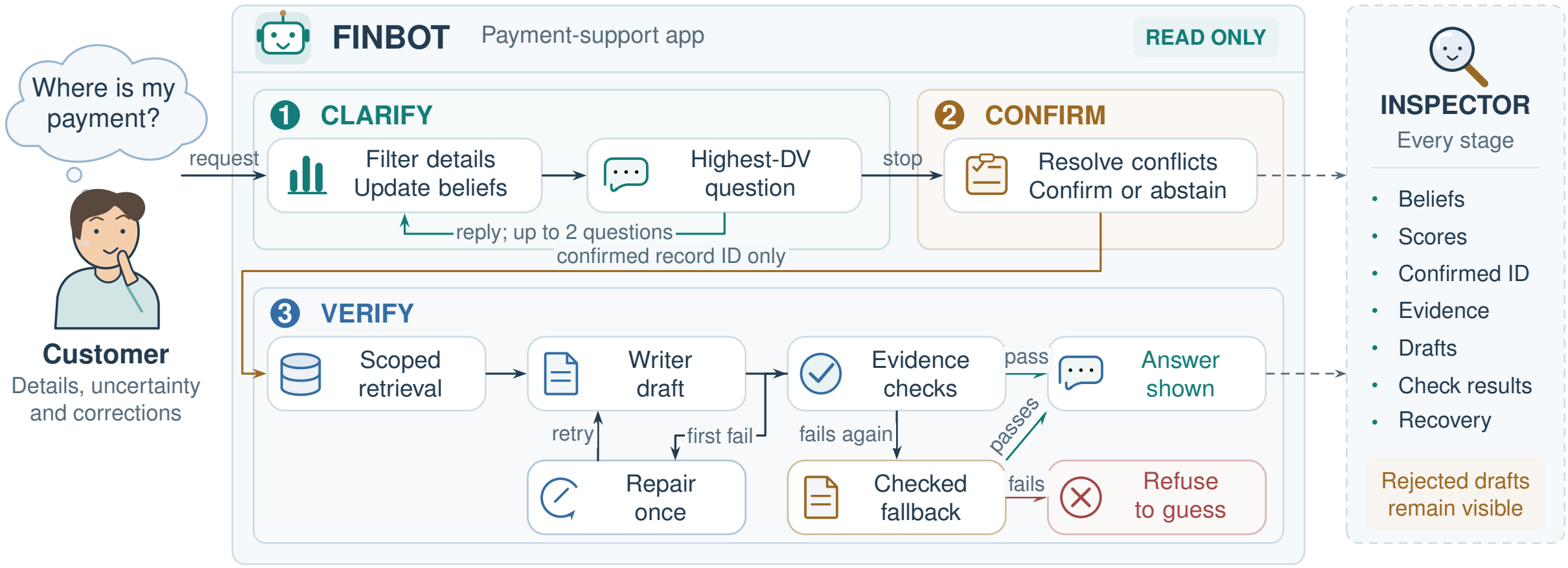}
\caption{Clarification filters supplied details and scores questions over matching records.
Conflicting selections require explicit resolution; customers may leave unresolved. Code restricts
answering to the confirmed ID and governs verification and recovery. Dashed links show inspector
observations. The separate inspector also supports direct comparisons and policy questions.}
\label{fig:architecture}
\end{figure*}

Bayesian question selection \citep{lindley1956measure} and evidence-based revision
\citep{gao-etal-2023-rarr} address parts of this; financial agents already use confirmation and
guardrails \citep{chua-etal-2025-banking}.
Finbot's central systems contribution is an inspectable architecture that preserves record
identity and claim roles across clarification, confirmation, scoped retrieval, verification and
recovery (Figure~\ref{fig:architecture}). Decision-value scoring over a belief state is the demo's
default question selector, not a claimed algorithmic advance: a most-distinct ordering beats it
with one question and matches it with two. The belief state earns its place by weighing hedged
details, carrying corrections and stopping early on expected selection error. We contribute:
\begin{itemize}\itemsep0pt
\item Financial-role claim checks and shared evidence requirements that guide writer
  instructions, verification and bounded recovery; traces show source fields, roles, missing
  information and scripted/live writer labels.
\item A controller whose belief state guides clarification, while explicit confirmation and
  conflict resolution govern retrieval for explanation.
\item Reproducible component simulations, controller replays, authored ablations and two small
  two-labeller audits; they do not establish customer accuracy.
\end{itemize}
Finbot is intended for developers and researchers building inspectable natural-language
interfaces over structured financial and customer-support records. The demo uses synthetic
records and policies and executes no payments, refunds or disputes.
The hosted demo is at \url{https://finbot-demo-20852907589.us-central1.run.app}.
Run \texttt{python -m finbot.demo} locally without a model account. Code, data, results and a 2.5-minute video are at
\url{https://github.com/soovadeep/finbot-demo} (MIT license).

\section{System}
\label{sec:system}

\paragraph{Interaction surfaces.} The conversation interprets messages and asks template-based
questions without an LLM, then answers about one confirmed record among four synthetic payments.
Eleven \texttt{/inspector} presets cover declines, unknown outcomes, comparisons, policies, financial roles and recovery.
Both share answering and verification; the conversation does not route arbitrary policy or
multi-record requests to these presets.

\subsection{Clarifying which record}

\paragraph{Candidates and beliefs.}
Let $b(h)$ be the belief that authorized record $h\in\mathcal{H}$ is intended, initially
$b_0(h)=1/|\mathcal{H}|$. Question attributes are kind (card or transfer), payee, date and amount.
The demo has four synthetic records; the component study uses up to eight. Clarification cannot
expand this authorized pool.

\paragraph{Questions and replies.}
A question $q$ offers distinct attribute values plus ``I'm not sure''. For reply $r$ and assumed
likelihood $\ell_q(r\mid h)$ (Appendix~\ref{app:repro}), Bayes' rule gives
\begin{equation}
b_{q,r}(h)=\frac{b(h)\,\ell_q(r\mid h)}{\sum_g b(g)\,\ell_q(r\mid g)}.
\label{eq:bayes}
\end{equation}
An unsure reply has the same likelihood for every record, leaving beliefs unchanged; a
zero-probability reply also retains the prior. Hedged details use lower assumed reliability.
These are response-model assumptions, not measured human accuracy.

\paragraph{Choosing the next question.}
Define $p_q(r)=\sum_h b(h)\ell_q(r\mid h)$ and entropy
$H(b)=-\sum_h b(h)\log_2 b(h)$. Information gain (IG) is the expected entropy reduction:
\begin{equation}
\mathrm{IG}(q)=H(b)-\sum_r p_q(r)\,H(b_{q,r}).
\label{eq:ig}
\end{equation}
With $e(b)=1-\max_h b(h)$, decision value (DV) is the expected reduction in maximum-belief
selection error, minus cost $c=0.02$:
\begin{equation}
\mathrm{DV}(q)=e(b)-\sum_r p_q(r)\,e(b_{q,r})-c.
\label{eq:dv}
\end{equation}
Sums include unsure replies. This hypothetical choice scores questions; it never selects a
record for answering. The component compares IG and DV with fixed order (kind, payee, date, amount),
amount-first and most-distinct, which asks the attribute offering the most distinct values among
the candidates (ties keep the fixed order). It stops at its question budget or 95\% belief;
IG and DV additionally require a positive score. The demo uses DV after renormalizing beliefs
over matching records, stopping at two questions, a single match, indistinguishable matches,
or no useful question.

\paragraph{Confirmation and corrections.}
The demo filters unhedged details while retaining alternatives for hedged details. A selection
conflicting with supplied type, payee, date, amount, currency or reference pauses retrieval for
explicit correction, reselection or abstention. Stale acknowledgements are rejected; changed
selections supersede prior answers. Conflict resolution is outside the two-question budget.
Consistent but mistaken confirmations remain possible: the system cannot infer hidden intent.

\subsection{Evidence retrieval, explanation and recovery}

The writer uses read-only tools to retrieve records and policy evidence. In the conversation,
the confirmed record is fetched again to answer the customer's own question; inspector requests
can name multiple records or ask about policy. Deterministic checks compare recognized amounts,
currencies, dates, statuses, causes and codes with eligible evidence. A failing draft gets one
repair, then a checked record-built answer or policy-only source quotation, then refusal. Fields such as merchant
descriptors cannot support factual claims.

\paragraph{Access boundaries.} The demo fixes a synthetic customer identity; it does not
authenticate real banking customers. Code binds the tenant and confirmed ID outside the model's
tool arguments. The conversation exposes only the confirmed record to the answering step;
other lookups return ``not found''. Mismatched returned IDs are rejected.

Regular expressions and phrase tables recognize supported claim formats, with limited negation
and hedge handling. Recognized values must occur in eligible evidence; recognized causes must
map to recorded reason codes. Policy statements retain complete cited passages; code questions
also require coverage of every requested code by passage text or section headings. Explicit
comparisons require each requested record and recognized requested fields, including in fallback.
Single-record coverage maps question topics to required statuses, amounts, timing and causes.
The same plan supplies writer instructions, checks and fallback; missing information must be
explicit. Refund arrival and payment settlement use separate fields. Fixed text explains schema
meaning and assistance limits. The latest record governs absence; known zero cannot be denied.
Coverage is finite: unrecognized wording, date formats and bare web addresses can evade extraction
(Appendix~\ref{app:inject}); omissions remain possible. A passing gate does not establish correctness.

\paragraph{Financial roles.} A recognized claim $c=(h,r,v)$ identifies a record, role and value.
The role maps to a schema path $\pi(r)$; support requires $v=D_h[\pi(r)]$ after normalization,
using the latest lookup $D_h$. Roles cover amounts and refund/settlement dates; actual settlement
dates require settled status. The inspector exposes the phrase, role, field
and verdict. Fallback states when actual dates are unavailable.
Example: a record holds a 200.00 payment and a 50.00 refund. The draft ``Refund amount: 200.00''
passes record binding (200.00 is on the record) but fails the role check (\texttt{refund.amount} is
50.00); the repair states 50.00 and passes (inspector: ``Payment is not refund''). An optional LLM
evidence review is off by default and not evaluated here.

\paragraph{Agent context.} Prompt admission reserves output capacity and rolls old turns into bounded
state; offline tests cover this (\texttt{docs/agent-design.md}).

\subsection{What the inspector shows}

Both interfaces expose evidence, drafts, rejection reasons and recovery; the conversation adds
beliefs and question scores, and the inspector adds policy excerpts and trace export. Fault
injection changes drafts while preserving evidence: a status or fee in the conversation; also
record ownership, financial roles or policy rules in the inspector. These test recovery, not natural LLM error rates.

\paragraph{Writers.} The default writer is scripted; optional Llama 3.3 70B through OpenRouter
uses the same scope, checks and recovery. Visitor/daily caps and provider failure trigger scripted
fallback; each answer is labelled scripted, newly generated, cached replay, or scripted after a cap or
failure. Clarification uses no LLM.

\section{Demonstration walkthrough}

\begin{figure*}[t]
\centering
\includegraphics[width=\textwidth]{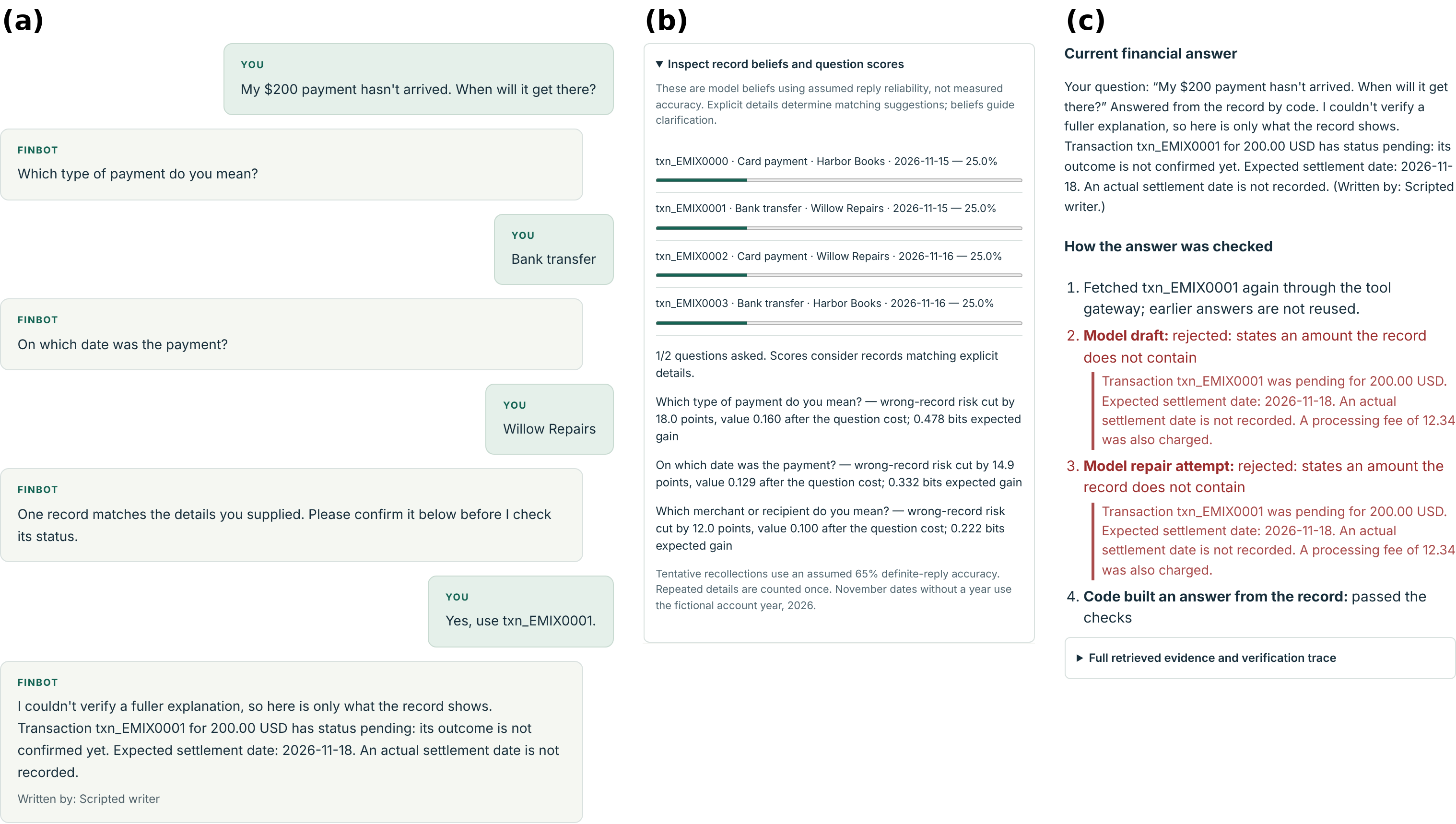}
\caption{Conversation walkthrough with ``wrong amount every time''. (a)~Two questions narrow
four \$200 payments before confirmation. (b)~Beliefs and question scores at the first question.
(c)~The customer's question is carried through; both drafts invent a fee, so the gate rejects them
and checks a record-built fallback that states the expected date. Each answer names its writer.}
\label{fig:screenshot}
\end{figure*}

Figure~\ref{fig:screenshot} shows clarification and record-built recovery. Correcting to
``card payment on November 16'' supersedes the answer and requires fresh confirmation.

\paragraph{Additional scenarios.} Inspector presets explain declines and unknown timeouts,
reject cross-record amounts and reversed cited policies, and distinguish a \$200 payment from
its \$50 refund. ``Expected is not actual'' shows a date-role rejection and repair. Failed
policy-only repairs trigger a cited passage fallback; absent documentation produces refusal. All policies are fictional.

\section{Evaluation}
\label{sec:eval}

\begin{table}[t]
\centering\small
\footnotesize
\setlength{\tabcolsep}{2.2pt}
\begin{tabular}{@{}lcccc@{\hspace{5pt}}ccc@{}}
\toprule
 & & \multicolumn{3}{c}{Haiku 5.5} & \multicolumn{3}{c}{Llama 3.3 70B} \\
\cmidrule(lr){3-5}\cmidrule(l){6-8}
Strategy & Qs & Top-1 & Corr. & Wr. & Top-1 & Corr. & Wr. \\
\midrule
None & 0 & 13 & 17 & 3.1 & 13 & 21 & 5.4 \\
Fixed, 1 & 1 & 19 & 24 & 2.2 & 19 & 30 & 5.8 \\
IG, 1 & 1 & 65 & 44 & 2.2 & 70 & 59 & 4.0 \\
DV, 1 & 1 & 69 & 47 & 2.2 & 74 & 63 & 4.0 \\
Amount-first, 1 & 1 & 63 & 42 & 2.2 & 75 & 63 & 4.0 \\
Most distinct, 1 & 1 & 72 & 50 & 2.2 & 77 & 66 & 4.0 \\
Fixed, 2 & 1.95 & 72 & 48 & 1.8 & 77 & 66 & 4.0 \\
IG, 2 & 1.95 & 73 & 46 & 1.8 & 78 & 66 & 4.0 \\
DV, 2 & 1.36 & 74 & 49 & 2.2 & 79 & 66 & 4.0 \\
\quad + fallback & 1.36 & n/a & 66 & 3.1 & n/a & 81 & 4.9 \\
Show all 8 & 0 & n/a & 60 & 1.8 & n/a & 81 & 3.6 \\
\bottomrule
\end{tabular}
\caption{Eight-record component study, 224 episodes (\%). Fixed: fixed order; IG: information gain;
DV: decision value. Suffix: question budget. Qs: Haiku mean (Llama DV2: 1.31). Top-1: target
ranked first. Corr./Wr.: correct/wrong confirmations from the two-record shortlist or full list.
None shows the first two records; fallback shows all eight after rejection. Heuristic rows are
exploratory cached comparisons. Correct-rate intervals are approximately $\pm$6 points.}
\label{tab:main}
\end{table}

\begin{table}[t]
\centering\small
\setlength{\tabcolsep}{3pt}
\begin{tabularx}{\columnwidth}{@{}p{.31\columnwidth}X@{}}
\toprule
Capability / demo & Evidence and boundary \\
\midrule
Clarification / conversation & 224 scenarios, two simulators and stronger baselines; component policy differs. \\
Confirmation, corrections / conversation & 60 controller replays; unit and browser checks include conflicts. Not customer accuracy. \\
Explanations, recovery / both views & 3,240 template mutation pairs with matched controls (Appendix~\ref{app:inject}); 128-draft live audit. Semantic errors remain. \\
Financial roles / inspector & 60 corruption pairs, 12 equal-value controls, replayed cached drafts; finite claim formats. \\
Requested fields / both & 30 unused public questions, paired pre/post-fix: 23/60 to 50/60 complete (AI); blind co-author 56/60 post-fix. \\
Record comparison / inspector & Scripted presets only; not evaluated here. \\
Policy, recovery / inspector & 30 new live Llama cases, paired configurations; two labellers. \\
Trace inspection / both views & Browser and accessibility checks (Appendix~\ref{app:ui}); no comprehension study. \\
\bottomrule
\end{tabularx}
\caption{Capability-to-evidence map. Inspector runs and mutations are authored synthetic
checks; public wording is paired with synthetic records. No customer usability study is complete.}
\label{tab:capabilities}
\end{table}

\subsection{Ambiguity benchmark}

We pair 224 Banking77 test questions \citep{casanueva-etal-2020-efficient} with synthetic card
payments or transfers and remove their added references. Same-customer distractors prefer the
target's payment kind and nearby dates (seed 2027). The main set has eight candidates with
unique amounts; a smaller set has two to five (mean 3.6). Restricting to the 192 questions
consistent with their paired records changes the original rates by at most three points
(Appendix~\ref{app:bench}). Section~\ref{sec:same} removes the unique-amount shortcut.

\subsection{Clarification component study}

This study uses \texttt{ClarificationSession}, which updates beliefs from option replies and
offers a ranked two-record shortlist. The deployed \texttt{Conversation} controller also parses
openings and free text, filters on unhedged details and exposes other records. Shared scoring
functions do not make these policies equivalent; the rates below describe the component only.
Separate controller replays use 15 authored event sequences and four fault settings, including
conflicting selections. All 60 meet expected behavior with a scripted writer and no model calls
(Appendix~\ref{app:ui}). These are regression checks, not customer accuracy.

A simulated customer sees its opening and memory note, then candidates only at confirmation.
Seeded personas are \emph{receipt} (40\%, all details), \emph{vague} (35\%, partial details) and
\emph{misremembers} (25\%, one confident wrong detail). Claude Haiku 5.5 and Llama 3.3 70B
Instruct answer options, including ``not sure''. Replies are cached per question and confirmation
list, giving selectors identical responses to identical requests. The original study made 9,501
model calls for under US\$5; added baseline replays make none.

We compare the five selectors, no questions (show the first two records), and showing every
candidate at once. An episode is \emph{correct}
when the customer confirms the target, \emph{wrong} when they confirm another record, and
\emph{unresolved} when they pick none. Rate intervals are 95\% Wilson intervals; the added
paired baseline comparisons use target-record-cluster bootstrap intervals.

\paragraph{Results.}
Against fixed order, DV reaches similar two-question correctness with fewer questions
(1.31--1.36 versus 1.95; Table~\ref{tab:main}).

Stronger untuned heuristics qualify this comparison. One most-distinct question yields
49.6\%/66.1\% correct confirmations (Haiku/Llama), versus 46.9\%/62.5\% for one DV question;
amount-first yields 41.5\%/63.4\%. The paired one-question gap favours most-distinct by
2.7/3.6 points (cluster-bootstrap 95\% intervals 0.5--5.2 and 0.9--7.0); with two questions
each, the two differ by at most 1.3 points and the intervals include zero. Most-distinct also
wins on the same-amount set (32.6\%/46.9\% versus 29.9\%/43.3\%, one question).
We keep the belief state because the controller needs more than an order: it weighs hedged
details, carries evidence through corrections, stops early and exposes inspectable probabilities.

Showing all records yields more correct confirmations than the shortlist (60\%/81\%). In a
component variant, offering the full list after shortlist rejection raises DV correctness to
66\%/81\%, with 49\%/30\% of customers seeing eight records, but wrong selections rise to
3.1\%/4.9\%. The comparison measures
simulated identification and question count; human effort and completion time remain unmeasured.

\paragraph{Without unique amounts.}
\label{sec:same}
Giving all candidates the target's amount and currency leaves only 99 distinguishable targets.
On these, two-question DV confirms 68\%/85\% correctly, versus 73\%/88\% for the full list
(Appendix~\ref{app:same}). The other 125 have indistinguishable twins. For these, the component
ranks a shortlist; the demo requests a reference or abstention.

Appendix~\ref{app:studies} reports reliability sensitivity, a Sonnet question chooser and
free-text replies, where parsing failures reduce confirmations to 42\%/51\%.

\subsection{Safety under wrong assumptions}
\label{sec:safety}

Acting on the top guess after the questions would pick the wrong record in 26\% (Haiku) and
21\% (Llama) of episodes; with confirmation, wrong selections are 2.2\% (0.9--5.1\%) and 4.0\%
(2.1--7.5\%), numerically higher than showing the whole list (1.8\% and 3.6\%). Overlapping
intervals do not establish equivalence or non-inferiority. With every customer confidently wrong
about one detail, top-guess selection is wrong in 42\%; confirmation yields 5.4\%/5.8\% wrong,
versus 4.5\%/7.6\% for the full list. Confirmation limits, but cannot eliminate, wrong-record risk.

\subsection{Answer checks}
\label{sec:inject}

\paragraph{Role ablation.} Six records and two wordings yield 60 value-preserving substitutions:
48 within-record, 12 cross-record. Detection is 0/60 without binding, 12/60 with record binding,
60/60 with financial roles; all 60 clean and 12 equal-value controls pass. These development
cases do not estimate accuracy; unrecognized date/refund phrasing still evades role extraction.

\paragraph{Template mutations.} On held-out template wordings, the record checks (without role
checks) reject 82.3\% (2,668/3,240) of corrupted drafts, but 570 clean controls already fail; among the 2,670 accepted controls,
mutations cause 78.6\% new rejections (Appendix~\ref{app:inject}). An earlier AI-assisted author audit
of 128 live Llama drafts found nine expected/actual-date errors, which informed the role checks that now
block them, and one omitted refund.

\paragraph{Live policy answers.} Llama 3.3 70B drafted answers for 30 new frozen cases (10 policy).
A request cache gives two configurations byte-identical drafts and repairs; they differ only in
a fallback that, after a policy-only draft and its repair both fail, quotes the best retrieved
passage, which must cover any requested code. It cuts refusals from 13 to 4 (two appropriate);
released answers rise from 17 to 26. Two labellers judged each of the 43 distinct answers
(drafts, repairs and fallbacks) for support and completeness: Claude, unblinded, and a co-author
blind to gate outcomes and to Claude's labels. They agree on 41/43; under either set no released
answer is wrong, and checks block 16 of 29 correct first drafts.

\paragraph{Public questions.} After a development set exposes omissions, we freeze code and sampling
before selecting 30 unused Banking77 train questions with prior-overlap screening. Two synthetic
record conditions yield 60 paired pre/post-fix tasks. Unblinded AI review (Codex) judges
23/60 versus 50/60 supported and complete (none regressed); a blind co-author
judges 56/60 post-fix answers so (52/60 agree, $\kappa$=0.37); the remaining incomplete
answers all pass the gate. These are scripted post-identification answers, not
live-writer or customer results. Both studies are small controlled audits, not external
validation; rubrics and label files are in Appendix~\ref{app:repro}.

\paragraph{Lessons.} Confirmation, not question choice, drives safety: every selector stays at
2--6\% wrong once customers confirm, versus 21--26\% for the top guess. Conservative checks need a
good fallback: they block about half of correct live drafts, yet no wrong live answer is released.

\section{Related work}

Prior work ranks clarification by usefulness \citep{rao-daume-iii-2018-learning}, studies search
\citep{aliannejadi2019asking,zhang2020evaluating}, selective questions \citep{kuhn2022clam},
information gain \citep{piriyakulkij2023active,li2025eliciting} and simulated users \citep{sekulic2022evaluating}.
Finbot targets authorized-record ambiguity and runtime checking of typed fields, complementing
attribution and atomic-fact metrics \citep{rashkin-etal-2023-measuring,min-etal-2023-factscore}.
Evidence-based revision \citep{gao-etal-2023-rarr}, programmable rails \citep{rebedea-etal-2023-nemo}
and control/data separation \citep{debenedetti2025camel} motivate its answering step.
Ryt AI executes banking operations with guardrails and confirmation
\citep{chua-etal-2025-banking}; Finbot is read-only and preserves record identity and claim roles.
Certainty distortions and omissions are also studied by \citet{nnamdi-etal-2026-epistemic}.
Finbot connects record confirmation to inspectable financial-role checks and recovery;
it introduces neither a Bayesian method nor a general semantic verifier.

\section{Limitations}

Simulators and synthetic records do not establish user performance. Banking77 questions were not
written for these records; assumed reliabilities and unique amounts favor amount questions.
Candidates are capped at eight. Component and controller
policies differ: confident-detail filters can exclude the target, and conflict checks cannot
infer intent. Public wording is adapted to synthetic records, and pre-fix labels are AI-only. Finite English checks
miss semantic errors; quotation does not prove relevance, nor context budgets long-chat recall.
Role-check evidence is the 60 authored substitutions and nine earlier-found live errors;
coverage of unseen human wording is unmeasured. The record-built fallback is a system component,
not an evaluated baseline: we have not tested an always-template system on the same tasks and
claim no advantage of model-written answers over field-based replies.
Live answers come from one writer. No external annotator, user, voice or
deployed-customer study is complete; Finbot is not validated for production financial support.

\section*{Acknowledgements}

Claude Haiku 5.5 and Llama 3.3 70B simulate customers, Claude Sonnet 5.5 is a baseline question
chooser and Llama 3.3 70B writes live answers. Claude and Codex labelled the audits unblinded (a
co-author blind); Claude Code and OpenAI Codex assisted coding, writing and the video.

\section*{Ethics and broader impact}

Wrong-record answers and unsupported financial claims can cause fees, repeated charges or missed
disputes. The demo uses synthetic data, involves no participants and executes no payments.
Deployment would need independent validation, consent and access controls.

\bibliography{references}

\clearpage
\appendix
\raggedbottom

\section{Reproducibility}
\label{app:repro}

Simulated customers: \texttt{claude-haiku-5-5} (Anthropic API, default sampling, no extended
thinking) and \texttt{meta-llama/llama-3.3-70b-instruct} (OpenRouter, temperature 0); chooser:
\texttt{claude-sonnet-5-5}; live writer: the Llama model (at most 400 output tokens). The benchmark
uses seed 2027 and personas are seeded by episode ID, so every selector meets the same customer.
Simulator responses are cached; prompt templates remain in the scripts. The added baseline
replay makes no model calls and reports missing coverage explicitly
(\texttt{scripts/naacl2027/}). Prompts, settings, the 192-episode comparison and the live-writer
audit are in \url{docs/papers/naacl-2027/evaluation/paper-details.md} in the repository. The two audits are
hash-frozen to the code that produced them; their replays run at the commits listed in
\url{evaluation/frozen-studies.md}.

\paragraph{Audit labels.} Both audits use one rubric: \emph{supported} means every claim agrees
with the record or policy in the right role; \emph{complete} means the answer gives what was asked
that the evidence contains, or says it is missing. Labels are never inferred from checker
verdicts. Blank forms, both label sets with provenance and agreement are in
\texttt{results/naacl2027/fresh-llm-v3/} and \texttt{public-wording-v2/}.

\paragraph{Response model.} In Eq.~\ref{eq:bayes}, attribute $k$ assigns likelihood $u_k$ to
uncertainty, $(1-u_k)a_k$ to the matching option, and divides the remainder uniformly among
wrong options. Settings $(a_k,u_k)$ are $(0.9,0.1)$ for kind, $(0.85,0.15)$ for date and
$(0.8,0.2)$ for payee and amount; hedged details use $a=0.65$. Replies across attributes are assumed
conditionally independent given the record. The demo rebuilds from a uniform prior with one
current value per attribute; corrections replace observations and repeats add no evidence.
These are model assumptions, not measured human reliabilities.

\section{Benchmark and personas}
\label{app:bench}

\paragraph{Naturally fitting.} A question is naturally fitting when nothing it says contradicts its
record. The 32 that fail ask about an ATM withdrawal (paired with a card payment) or a pending
payment (paired with a record that is not pending). On the 192 that fit, every strategy's results
are within three points of those on all 224 (decision value with two questions: 51.6\% and 67.7\%
correct, against 48.7\% and 66.1\%).

\paragraph{Personas.} Haiku leaves 54\% of vague customers unresolved even with all eight records
shown (Llama 1\%): it declines to pick when its note does not settle the choice, as its prompt asks.
For the same reason vague Haiku customers resolve more with the fallback than with the full list
(57\% against 46\%): after the questions the short list holds fewer records that fit their note.
Both simulators fail on customers who misremember, who reject short lists that do not match their
memory.

\begin{table}[H]
\centering\small
\setlength{\tabcolsep}{2.5pt}
\resizebox{\columnwidth}{!}{%
\begin{tabular}{@{}llcccc@{}}
\toprule
 & Persona ($n$) & DV1 top-1 & DV2 correct & + fallback C/W/U & All 8 C/W/U \\
\midrule
\multicolumn{6}{@{}l}{\emph{Haiku 5.5 customers}} \\
 & Receipt (89) & 99 & 80 & 98/1/1 & 96/1/3 \\
 & Vague (72) & 43 & 46 & 57/0/43 & 46/0/54 \\
 & Misremembers (63) & 57 & 8 & 30/10/60 & 27/5/68 \\
\midrule
\multicolumn{6}{@{}l}{\emph{Llama 3.3 70B customers}} \\
 & Receipt (89) & 99 & 96 & 100/0/0 & 98/0/2 \\
 & Vague (72) & 58 & 76 & 93/6/1 & 97/1/1 \\
 & Misremembers (63) & 57 & 13 & 40/11/49 & 40/11/49 \\
\bottomrule
\end{tabular}}
\caption{Results by persona, eight records (\%). C/W/U: correct, wrong and unresolved.}
\label{tab:persona}
\end{table}

\section{Sensitivity, chooser and free text}
\label{app:studies}

\begin{table}[H]
\centering\small
\setlength{\tabcolsep}{2.6pt}
\resizebox{\columnwidth}{!}{%
\begin{tabular}{@{}llcccc@{}}
\toprule
& Setting & Qs & Top-1 & Correct & Wrong \\
\midrule
\multicolumn{6}{@{}l}{\emph{Haiku 5.5 customers}} \\
 & Default ($c = 0.02$) & 1.36 & 73.7 & 48.7 & 2.2 \\
 & One reliability, 0.6--0.99 & 1.21 & 73.2 & 49.6 & 2.2 \\
 & 20 random draws & 1.17--1.75 & n/a & 40--50 & 2.2--2.7 \\
 & $c = 0$ / $c = 0.1$ & 1.50 / 1.17 & 72.8 / 73.2 & 49.1 / 49.6 & 2.7 / 2.2 \\
 & Sonnet 5.5 chooses & 1.29 & 62.9 & 42.9 & 2.2 \\
 & Free-text replies & 1.51 & 62.1 & 42.0 & 2.2 \\
\midrule
\multicolumn{6}{@{}l}{\emph{Llama 3.3 70B customers}} \\
 & Default ($c = 0.02$) & 1.31 & 78.6 & 66.1 & 4.0 \\
 & One reliability, 0.6--0.99 & 1.16 & 78.1 & 66.1 & 4.0 \\
 & 20 random draws & 1.11--1.75 & n/a & 63--67 & 4.0--4.9 \\
 & $c = 0$ / $c = 0.1$ & 1.45 / 1.12 & 77.7 / 78.1 & 66.1 / 66.1 & 4.0 / 4.0 \\
 & Sonnet 5.5 chooses & 1.18 & 74.1 & 62.9 & 4.0 \\
 & Free-text replies & 1.48 & 61.6 & 50.9 & 4.9 \\
\bottomrule
\end{tabular}}
\caption{Decision value with a budget of two, eight records, mixed personas (\%). ``One
reliability'' sets every attribute's $a_k$ to the same value; every value from 0.6 to 0.99 gives
the same row. Random draws set each $a_k$ independently between 0.6 and 0.99.}
\label{tab:studies}
\end{table}

\section{Same-amount slice}
\label{app:same}

With two questions, fixed order puts the target first slightly more often than decision value
(49\% against 47\%) because it always asks both questions (1.38 on average), while decision value
stops when a second question cannot change the most likely record (1.16); confirmations are equal.
With 95\% Wilson intervals, one question puts the target first in 46\% (39--52) of all episodes
with decision value and 16\% (12--21) with fixed order, and in 85\% (76--91) against 18\% (12--27)
of the 99 identifiable ones. On those 99, two decision-value questions confirm the target in 68\%
(58--76, Haiku) and 85\% (76--91, Llama); the full list gives 73\% (63--81) and 88\% (80--93).

\begin{table}[H]
\centering\small
\setlength{\tabcolsep}{2.5pt}
\resizebox{\columnwidth}{!}{%
\begin{tabular}{@{}lccccccc@{}}
\toprule
 & & \multicolumn{2}{c}{Top-1} & \multicolumn{2}{c}{Haiku C/W} & \multicolumn{2}{c}{Llama C/W} \\
\cmidrule(lr){3-4}\cmidrule(lr){5-6}\cmidrule(l){7-8}
Strategy & Qs & All & Ident. & All & Ident. & All & Ident. \\
\midrule
No questions & 0 & 13.4 & 16.2 & 12/15 & 20/2 & 17/29 & 24/2 \\
Fixed order, 1 & 0.95/1 & 16.1 & 18.2 & 14/8 & 24/3 & 22/31 & 28/2 \\
Info.\ gain, 1 & 0.95/1 & 39.3 & 70.7 & 28/8 & 56/2 & 39/31 & 68/2 \\
Decision value, 1 & 0.95/1 & 45.5 & 84.8 & 30/8 & 61/2 & 43/31 & 77/2 \\
Most distinct, 1 & 0.95/1 & 49.1 & 92.9 & 33/8 & 67/2 & 47/31 & 85/2 \\
Fixed order, 2 & 1.38/1.97 & 49.1 & 92.9 & 31/8 & 63/3 & 47/31 & 85/2 \\
Decision value, 2 & 1.16/1.46 & 47.3 & 88.9 & 33/8 & 68/3 & 47/31 & 85/2 \\
Show all 8 & 0 & n/a & n/a & 35/7 & 73/0 & 50/32 & 88/2 \\
\bottomrule
\end{tabular}}
\caption{Eight records that all share the target's amount and currency (\%). All: 224 episodes;
Ident.: the 99 whose target differs from every other candidate in kind, payee or date. Qs: mean
questions (all/ident.). Top-1 from Haiku's replies (Llama's differ by at most one point). C/W:
correct and wrong confirmations.}
\label{tab:same}
\end{table}

\section{Answer checks with matched controls}
\label{app:inject}

\begin{table}[H]
\centering\small
\setlength{\tabcolsep}{3pt}
\begin{tabular}{@{}lrrr@{}}
\toprule
Mutation & Pairs & Clean blocked & New rejection \\
\midrule
Amount & 448 & 76 & 372/372 \\
Added fee & 448 & 76 & 372/372 \\
Currency & 224 & 38 & 186/186 \\
Record ID & 224 & 38 & 186/186 \\
Status & 552 & 114 & 394/438 \\
Date & 448 & 76 & 172/372 \\
Reason code & 224 & 38 & 58/186 \\
Bare link & 224 & 38 & 0/186 \\
Mixed wording & 448 & 76 & 358/372 \\
\midrule
Total & 3,240 & 570 & 2,098/2,670 \\
\bottomrule
\end{tabular}
\caption{Post-hoc paired analysis of the existing held-out wordings. New rejection: corrupted
blocked / clean accepted, conditional on an accepted control. All 570 rejected controls remain
rejected after mutation. Thus raw corrupted-draft rejection is 2,668/3,240 (82.3\%), while
mutation-triggered rejection among accepted controls is 78.6\%. Controls repeat across mutation
families; these are not independent samples. The separate 1,120-draft clean set has 114 false
rejections, all from ``has not settled yet'' about pending transfers.}
\label{tab:inject}
\end{table}

The paired report retains texts, verdicts and reasons. All 38 bare-link rejections predate
the mutation; no added link changes an accepted control's verdict. Reporting control rejection
alongside conditional sensitivity exposes false positives. These tests do not establish semantic completeness.

\section{Interface checks}
\label{app:ui}

Chromium passes 18 inspector checks across 320--1440-pixel layouts. axe-core 4.10.3 detects
no WCAG 2.1 A/AA violations in 14 states; eight later conflict-resolution states at 390 and
1280 pixels likewise show no violations or browser errors. These are not comprehensive
accessibility or comprehension studies.

Fifteen scripts under four fault settings give 60 expected runs: 32 correct, four wrong,
16 unresolved (eight with absent targets), eight awaiting clarification. Twenty have no draft
to corrupt; four paused runs supersede an earlier answer. Selection and field checks are
separate from verification and do not establish semantic accuracy.

\end{document}